\documentclass[12pt,a4paper]{article}
\usepackage{amsmath,amssymb,amsthm} 

\newtheorem{prop}{Proposition}

\theoremstyle{remark}
\newtheorem{remark}{Remark}
\newcommand{\beq}{\begin{equation}}
\newcommand{\eeq}{\end{equation}}
\newcommand{\beqnn}{\begin{equation*}}
\newcommand{\eeqnn}{\end{equation*}}
\newcommand{\rd}{\partial}

\newcommand{\CC}{\mathbb{C}}
\newcommand{\PP}{\mathbb{P}}

\newcommand{\ZZ}{\mathbb{Z}}
\newcommand{\frakL}{\mathfrak{L}}
\newcommand{\calA}{\mathcal{A}}

\newcommand{\calE}{\mathcal{E}}
\newcommand{\calL}{\mathcal{L}}
\newcommand{\bst}{\boldsymbol{t}}
\newcommand{\bsT}{\boldsymbol{T}}

\begin{document}

\title{Dressing operators in equivariant Gromov-Witten theory 
of $\mathbb{CP}^1$}
\author{Kanehisa Takasaki\thanks{E-mail: takasaki@math.kindai.ac.jp}\\
{\normalsize Department of Mathematics, Kindai University}\\ 
{\normalsize 3-4-1 Kowakae, Higashi-Osaka, Osaka 577-8502, Japan}}
\date{}
\maketitle

\begin{abstract}
Okounkov and Pandharipande proved that the equivariant 
Toda hierarchy governs the equivariant Gromov-Witten theory 
of $\mathbb{CP}^1$.  A technical clue of their method is a pair 
of dressing operators on the Fock space of 2D charged free fermion 
fields.  We reformulate these operators as difference operators 
in the Lax formalism of the 2D Toda hierarchy.  This leads 
to a new explanation to the question of why the equivariant 
Toda hierarchy emerges in the equivariant Gromov-Witten theory 
of $\mathbb{CP}^1$. Moreover, the non-equivariant limit 
of these operators turns out to capture the integrable structure 
of the non-equivariant Gromov-Witten theory correctly.  
\end{abstract}

\begin{flushleft}
\end{flushleft}

\newpage

\section{Introduction}

The Gromov-Witten theory of the Riemann sphere $\CC\PP^1$ 
is known to be related to integrable hierarchies 
of the Toda type.  Such a link was first observed 
by physicists employing random matrix models \cite{EY94,EHY95}.  
This discovery was enhanced to a mathematical statement 
called the Toda conjecture 
\cite{Pandharipande99,Okounkov00,Getzler01}.  
The Toda conjecture was proved by several methods 
\cite{OP02a,OP02b,Zhang02,DZ03,Milanov0605} 
and generalized to $\CC\PP^1$ with orbifold points 
\cite{MT0607,CvdL13,MST14,CM1910}.  
The relevant integrable hierarchies are 
the 1D Toda hierarchy, the bigraded Toda hierarchy 
\cite{CDZ04,Carlet06} and a kind of 
the Kac-Wakimoto hierarchies \cite{CM1909}. 
Another direction of generalization is the equivariant 
Gromov-Witten theory of $\CC\PP^1$ 
\cite{OP02b,Milanov0508,MT0707,Johnson09}. 
The integrable hierarchies emerging therein are 
the equivariant Toda hierarchy \cite{Getzler04} 
and its bigraded version \cite{MT0707}. 

The most exotic among these integrable hierarchies 
will be the equivariant Toda hierarchy.  
This integrable hierarchy, like the 1D Toda hierarchy, 
is a reduction of the 2D Toda hierarchy.  
Okounkov and Pandharipande \cite{OP02b} 
encode the Gromov-Witten invariants 
into the vacuum expectation value of an operator product 
on the fermionic Fock space of 2D charged free fermion fields. 
This fermionic expression can be converted 
into a tau function of the 2D Toda hierarchy, 
and eventually turns out to be a tau function 
of the equivariant Toda hierarchy. 
Milanov \cite{Milanov0508} starts from a bosonic expression 
of the Gromov-Witten invariants 
in the Givental theory \cite{Givental01}, 
and derives the Hirota bilinear equations 
of the equivariant Toda hierarchy.  
These results are further generalized to $\CC\PP^1$ 
with two orbifold points \cite{MT0707,Johnson09}. 

This paper reconsiders Okounkov and Pandharipande's 
method for the equivariant Gromov-Witten theory 
from the perspective of the Lax formalism 
of the 2D Toda hierarchy.  
A central role in their method is played by 
what they call dressing operators.
\footnote{This terminology is somewhat confusing, 
because dressing operators of a different kind are already 
used in the Lax formalism of the 2D Toda hierarchy.}
These operators on the fermionic Fock space are used 
to convert the fermionic expression of the Gromov-Witten 
invariants into a standard expression of tau functions 
of the 2D Toda hierarchy.  It is, however, not very clear 
from this expression that the tau function is indeed 
a tau function of the equivariant Toda hierarchy.  
We reformulate Okounkov and Pandharipande's dressing operators 
as difference operators like the Lax operators 
of the 2D Toda hierarchy.  This enables us to explain 
the relation to the equivariant Toda hierarchy 
in a more direct manner.  Throughout this paper, 
the ordinary equivariant Toda hierarchy and its orbifold 
generalizations are treated on an equal footing.

\section{Equivariant Toda hierarchy as reduction of 2D Toda hierarchy}

Let us recall the Lax formalism of the 2D Toda hierarchy 
(see the recent review \cite{Takasaki18}). 
$\bst = \{t_k\}_{k=1}^\infty$ and 
$\bar{\bst}= \{\bar{t}_k\}_{k=1}^\infty$ 
are the two sets of time variables. 
$s$ is the spatial coordinate, which is understood 
to be a continuous variable throughout this paper.  
$\Lambda$ denotes the shift operator 
\[
  \Lambda = e^{\rd_s}, \quad \rd_s = \rd/\rd s. 
\]
The Lax operators $L,\bar{L}$ of the 2D Toda hierarchy 
are difference (or pseudo-difference) operators of the form 
\[
\begin{gathered}
  L = \Lambda + \sum_{n=1}^\infty u_n\Lambda^{1-n},\quad 
  \bar{L}^{-1} = \sum_{n=0}^\infty\bar{u}_n\Lambda^{n-1},\\
  u_n = u_n(s,\bst,\bar{\bst}),\quad 
  \bar{u}_n =\bar{u}_n(s,\bst,\bar{\bst}),\quad 
  \bar{u}_0(s,\bst,\bar{\bst}) \not= 0, 
\end{gathered}
\]
and satisfy the Lax equations
\[
\begin{gathered}
  \frac{\rd L}{\rd t_k} = [B_k,L],\quad 
  \frac{\rd L}{\rd\bar{t}_k} = [\bar{B}_k,L],\\
  \frac{\rd \bar{L}}{\rd t_k} = [B_k,\bar{L}],\quad 
  \frac{\rd \bar{L}}{\rd\bar{t}_k} = [\bar{B}_k,\bar{L}],
\end{gathered}
\]
where 
\[
  B_k = (L^k)_{\geq 0},\quad \bar{B}_k = (\bar{L}^{-k})_{<0}. 
\]
$(\quad)_{\ge 0}$ and $(\quad)_{<0}$ denote the projection 
onto the non-negative and negative powers of $\Lambda$: 
\[
  \left(\sum_{n\in\ZZ}a_n\Lambda^n\right)_{\ge 0} 
    = \sum_{n\geq 0}a_n\Lambda^n,\quad 
  \left(\sum_{n\in\ZZ}a_n\Lambda^n\right)_{<0} 
    = \sum_{n<0}a_n\Lambda^n. 
\]
Let us mention that these equations should be formulated 
in the $\hbar$-dependent form \cite{TT95} to accommodate 
the $\hbar$-expansion (i.e., genus expansion) 
of the Gromov-Witten theory. To avoid notational complexity, 
however, we dare not to consider the $\hbar$-dependent form.  
We can move to the $\hbar$-dependent formulation, 
at least formally, by rescaling the variables 
as $t_k \to t_k/\hbar$, $\bar{t}_k\to \bar{t}_k/\hbar$ 
and $s \to s/\hbar$. 

The Lax operators can be expressed in a dressed form as 
\[
  L = W\Lambda W^{-1},\quad 
  \bar{L} = \bar{W}\Lambda\bar{W}^{-1} 
\]
with the dressing operators 
\[
\begin{gathered}
    W = 1 + \sum_{n=1}^\infty w_n\Lambda^{-n},\quad 
  \bar{W} = \sum_{n=0}^\infty\bar{w}_n\Lambda^n,\\
  w_n = w_n(s,\bst,\bar{\bst}),\quad 
  \bar{w}_n = \bar{w}_n(s,\bst,\bar{\bst}),\quad 
  \bar{w}_0(s,\bst,\bar{\bst}) \not= 0.
\end{gathered}
\]
The dressing operators $W,\bar{W}$ satisfy 
the Sato equations 
\[
\begin{gathered}
  \frac{\rd W}{\rd t_k} = B_kW - W\Lambda^k,\quad 
  \frac{\rd W}{\rd\bar{t}_k} = \bar{B}_kW,\\
  \frac{\rd\bar{W}}{\rd t_k} = B_k\bar{W},\quad 
  \frac{\rd\bar{W}}{\rd\bar{t}_k} = \bar{B}_k\bar{W} - W\Lambda^{-k}. 
\end{gathered}
\]
The logarithm of $L,\bar{L}$ can be defined 
with the dressing operators as 
\[
\begin{gathered}
  \log L = W\log\Lambda W^{-1} 
    = \rd_s - \frac{\rd W}{\rd s}W^{-1},\\
  \log\bar{L} = \bar{W}\log\Lambda\bar{W}^{-1}
    = \rd_s - \frac{\rd\bar{W}}{\rd s}\bar{W}^{-1}. 
\end{gathered}
\]
Note that $\log\Lambda = \rd_s$. 

Let $a,b$ be positive integers.  They are related 
to the orders of two orbifold points of $\CC\PP^1$.  
The case of $a = b = 1$ amounts to the ordinary $\CC\PP^1$. 
The equivariant Toda hierarchy of type $(a,b)$ 
can be obtained by imposing the reduction condition 
\cite{Getzler01,MT0707}
\beq
  L^a - \nu\log L = \bar{L}^{-b} - \nu\log\bar{L} - \nu\log Q, 
  \label{LLbar-rel}
\eeq
where $\nu$ and $Q$ are parameters of the reduction. 
$\nu$ is called the equivariant parameter.  
In the non-equivariant limit as $\nu \to 0$, 
the reduced system turns into the bigraded Toda hierarchy 
of type $(a,b)$ \cite{Carlet06}. 
$Q$ is related to the particular solution 
that we shall consider later on. 

(\ref{LLbar-rel}) implies that both sides 
become an operator of the form 
\beq
  \frakL = B_a + \bar{B}_b - \nu\log\Lambda. 
\eeq
The Lax equations of the 2D Toda hierarchy turn into 
the Lax equations 
\[
  \frac{\rd\frakL}{\rd t_k} = [B_k,\frakL],\quad 
  \frac{\rd\frakL}{\rd\bar{t}_k} = [\bar{B}_k,\frakL]
\]
for this reduced Lax operator.  
In the language of tau functions, (\ref{LLbar-rel}) 
amounts to the condition that the tau function 
depends on $s$ in the particular form 
\beq
  \tau = Q^{s^2/2}
        f(\{t_k + \delta_{ka}s/\nu,\bar{t}_k  + \delta_{kb}s/\nu\}_{k=1}^\infty). 
\eeq

\section{Dressing operators $V,\bar{V}$}

Okounkov and Pandharipande's dressing operators $V,\bar{V}$ 
\cite{OP02b} 
\footnote{These operators are denoted by $W,W^*$ 
in Okounkov and Pandharipande's notation.  
We have changed the notation to avoid confusion 
with the aforementioned dressing operators $W,\bar{W}$ 
for the Lax operators.}
are elements of the $GL(\infty)$ group acting 
on the Fock space of 2D charged free fermion fields. 
The tau function of the equivariant Gromov-Witten theory, 
generalized to $\CC\PP^1$ with two orbifold points 
of orders $a,b$ \cite{Johnson09}, can be thereby 
expressed as 
\beq
  \tau = \langle s|\exp\left(\sum_{k=1}^\infty t_kJ_k\right)
         g\exp\left(- \sum_{k=1}^\infty\bar{t}_kJ_{-k}\right)|s\rangle,
  \label{tau}
\eeq
where 
\beq
  g = V^{-1}e^{J_a/a}Q^{L_0}e^{J_{-b}/b}\bar{V}^{-1}. 
  \label{g}
\eeq
$J_k$'s are the generators of the $U(1)$ current algebra 
of the free fermion system, and $L_0$ is the zero mode 
of the Virasoro algebra therein.  

Okounkov and Pandharipande's construction of $V,\bar{V}$ 
relies on the correspondence between fermionic operators 
and $\ZZ\times\ZZ$ matrices.  Those matrices can be 
further represented by difference operators 
on the discrete space $\ZZ$, e.g., 
\beq
  J_k \longleftrightarrow \Lambda^k,\quad 
  L_0 \longleftrightarrow H = s - 1/2. 
  \label{fermion-dop}
\eeq
Actually, $\Lambda^k$ and $H$ are meaningful 
in the continuous space as well.  We shall redefine 
$V$ and $\bar{V}$ as well to be difference operators 
on the continuous space.  

In our reformulation, $V$ and $\bar{V}$ are difference operators 
of the form 
\footnote{This implies that their fermionic counterparts 
leave invariant the ground state of the charge-$s$ sector 
in the Fock space as $\langle s|V = \langle s|$ 
and $\bar{V}|s\rangle = |s\rangle$.}
\[
\begin{gathered}
  V =  1 + \sum_{n=1}^\infty v_n\Lambda^{-n},\quad 
  \bar{V} = 1 + \sum_{n=1}^\infty\bar{v}_n\Lambda^n,\\
  v_n = v_n(s),\quad \bar{v}_n = \bar{v}_n(s), 
\end{gathered}
\]
and satisfy the following intertwining relations: 
\begin{align}
  (\Lambda^a + H - \nu\log\Lambda)V 
    &= V(\Lambda^a - \nu\log\Lambda), \label{V-rel}\\
  \bar{V}(\Lambda^{-b} + H - \nu\log\Lambda) 
    &= (\Lambda^{-b} - \nu\log\Lambda)\bar{V}. \label{Vbar-rel}
\end{align}
These operators can be constructed by power series expansion 
with respect to $\nu$, see Section 5. 
We shall show in the next section that (\ref{V-rel}) 
and (\ref{Vbar-rel}) lead correctly to a solution 
of the reduction condition (\ref{LLbar-rel}). 

As further evidence for the validity of our reformation, 
let us mention that (\ref{V-rel}) and (\ref{Vbar-rel}) 
in the case of $a = b = 1$ are consistent with 
Okounkov and Pandharipande's intertwining relations 
between two operators $\calA(z,w)$ and $\tilde{\calA}(z,w)$ 
on the fermionic Fock space.  The counterparts 
of these operators on the difference operator side 
of the correspondence (\ref{fermion-dop}) are 
\[
\begin{aligned}
  \calA(z,w) &= \left(\frac{\zeta(w)}{w}\right)^z
     \sum_{k\in\ZZ}\frac{\zeta(w)^k}{(1+z)_k}\calE_k(w),\\
  \tilde{\calA}(z,w) &= \sum_{k\in\ZZ}\frac{w^k}{(1+z)_k}\Lambda^k,
\end{aligned}
\]
where 
\[
  \calE_k(z) = e^{z(H + k/2)}\Lambda^k,\quad 
  \zeta(z) = e^{z/2} - e^{-z/2},\quad 
  (1+z)_k = \frac{\Gamma(1+z+k)}{\Gamma(1+z)}. 
\]
For matching with Okounkov and Pandharipande's notation, 
let us introduce the new parameter $u = 1/\nu$ 
and rewrite (\ref{V-rel}), specialized to $a = b = 1$, as 
\beq
  (u(\Lambda + H) - \log\Lambda)V = V(u\Lambda - \log\Lambda). 
  \label{V-rel2}
\eeq

\begin{prop}
(\ref{V-rel2}) implies the intertwining relations 
\beq
  \calA(m,mu)V = V\tilde{\calA}(m,mu),\quad m = 1,2,\ldots. 
  \label{AV=VAtilde}
\eeq
\end{prop}

\proof
Exponentiating both sides of (\ref{V-rel2}), 
we have the identity 
\[
  e^{m(u\Lambda + uH - \log\Lambda)}V = Ve^{m(u\Lambda - \log\Lambda)}.
\]
The exponential on the right side boils down to 
\[
  e^{m(u\Lambda - \log\Lambda)} = e^{mu\Lambda}\Lambda^{-m}. 
\]
Since $[\log\Lambda,H] = 1$, we can use 
the Baker-Campbell-Hausdorff formula to compute 
the exponential on the left side as 
\[
\begin{aligned}
  e^{m(u\Lambda + uH - \log\Lambda)}
  &= e^\Lambda e^{m(uH - \log\Lambda)}e^{-\Lambda}\\
  &= e^\Lambda e^{-mu^2/2}e^{muH}\Lambda^{-m}e^{-\Lambda}\\
  &= e^\Lambda\calE_{-m}(mu)e^{-\Lambda}. 
\end{aligned}
\]
Therefore 
\[
  e^\Lambda\calE_{-m}(mu)e^{-\Lambda}V= Ve^{mu\Lambda}\Lambda^{-m}.
\]
Since 
\[
\begin{aligned}
  \calA(m,mu) &= \frac{m!}{(mu)^m}e^\Lambda\calE_{-m}(mu)e^{-\Lambda},\\
  \tilde{\calA}(m,mu) &= \frac{m!}{(mu)^m}e^{mu\Lambda}\Lambda^{-m},
\end{aligned}
\]
we find that (\ref{AV=VAtilde}) holds.  
\qed

(\ref{AV=VAtilde}) takes exactly the same form as the one 
presented by Okounkov and Pandharipande \cite{OP02b}.   
We can derive a similar intertwining relation 
for $\bar{V}$ from (\ref{Vbar-rel}) in much the same way. 
Thus Okounkov and Pandharipande's intertwining relations 
can be, at least partially, recovered in our reformulation 
of the dressing operators $V,\bar{V}$.

\section{Algebraic relation of Lax operators $L,\bar{L}$}

The dressing operators $W,\bar{W}$ 
of the tau function (\ref{tau}) can be captured 
by the factorization problem 
(see the review \cite{Takasaki18}) 
\beq
  \exp\left(\sum_{k=1}^\infty t_k\Lambda^k\right)U
  \exp\left(- \sum_{k=1}^\infty\bar{t}_k\Lambda^{-k}\right) 
  = W^{-1}\bar{W}, 
  \label{factorization}
\eeq
where $U$ is the difference operator 
\beq
  U = V^{-1}e^{\Lambda^a/a}Q^He^{\Lambda^{-b}/b}\bar{V}^{-1}
  \label{U}
\eeq
that corresponds to the operator (\ref{g}) 
on the fermionic Fock space.  This operator satisfies 
the following intertwining relation, which implies 
the reduction condition (\ref{LLbar-rel}) 
to the equivariant Toda hierarchy. 

\begin{prop}
\beq
  (\Lambda^a - \nu\log\Lambda)U 
  = U(\Lambda^{-b} - \nu\log\Lambda - \nu\log Q). 
  \label{..U=U..}
\eeq
\end{prop}

\proof
We use (\ref{V-rel}) and the operator identities 
\[
\begin{aligned}
  e^{\Lambda^a/a}He^{-\Lambda^a/a} &= \Lambda^a + H,\\
  e^{-\Lambda^{-b}/b}He^{\Lambda^{-b}/b} &= \Lambda^b + H,\\
  \log\Lambda\,Q^H &= Q^H(\log\Lambda + \log Q)
\end{aligned}
\]
to derive (\ref{..U=U..}) from (\ref{U}) as 
\[
\begin{aligned}
  (\Lambda^a - \nu\log\Lambda)U 
  &= V^{-1}(\Lambda^a + H - \nu\log\Lambda)
     e^{\Lambda^a/a}Q^He^{\Lambda^{-b}/b}\bar{V}^{-1} \\
  &= V^{-1}e^{\Lambda^a/a}(H - \nu\log\Lambda)
     Q^He^{\Lambda^{-b}/b}\bar{V}^{-1} \\
  &= V^{-1}e^{\Lambda^a/a}Q^H(H - \nu\log\Lambda - \nu\log Q)
     e^{\Lambda^{-b}/b}\bar{V}^{-1} \\
  &= V^{-1}e^{\Lambda^a/a}Q^He^{\Lambda^{-b}/b}
     (\Lambda^{-b} + H - \nu\log\Lambda - \nu\log Q)\bar{V}^{-1}\\
  &= U(\Lambda^{-b} - \nu\log\Lambda - \nu\log Q). 
\end{aligned}
\]
\qed

\begin{prop}
The Lax operators obtained from the solution 
of the factorization problem (\ref{factorization}) 
satisfy the reduction condition (\ref{LLbar-rel}).
\end{prop}

\proof
Let us rewrite (\ref{factorization}) as 
\[
  U = \exp\left(- \sum_{k=1}^\infty t_k\Lambda^k\right)
      W^{-1}\bar{W}\exp\left(\sum_{k=1}^\infty\bar{t}_k\Lambda^{-k}\right) 
\]
and plug it into (\ref{..U=U..}). 
After some algebra, we find that 
\[
  W(\Lambda^a - \nu\log\Lambda)W^{-1}
  = \bar{W}(\Lambda^{-b} - \nu\log\Lambda - \nu\log Q)\bar{W}^{-1}.
\]
This is nothing but (\ref{LLbar-rel}). 
\qed

\section{Construction of $V,\bar{V}$}

We construct the dressing operators $V,\bar{V}$ 
by the power series expansion
\beq
  V = \sum_{k=0}^\infty\nu^kV_k, \quad 
  \bar{V} = \sum_{k=0}^\infty\nu^k\bar{V}_k
  \label{nu-expansion}
\eeq
with respect to $\nu$.  As it turns out below, 
$V_k$'s become difference operators of the form 
\beq
  V_0 = 1 + \sum_{n=1}^\infty v_{0n}\Lambda^{-n},\quad 
  V_k = \sum_{n=ka}^\infty v_{kn}\Lambda^{-n},\quad k\geq 1. 
  \label{V0Vk}
\eeq
$\bar{V}_k$'s, too, take a similar form.  
Since $\bar{V}$ can be obtained from the formal adjoint 
(or transpose) $V^*$ of $V$ as 
\beq
  \bar{V} = V^*|_{\nu\to -\nu,\,a\to b}, 
\eeq
we present the construction of $V$ only.  

(\ref{V-rel}) splits into the following set of equations 
for $V_k$'s: 
\begin{align}
  (\Lambda^a + H)V_0 &= V_0\Lambda^a, \label{V0-eq}\\
  (\Lambda^a + H)V_k - \log\Lambda V_{k-1}
    &= V_k\Lambda^k - V_{k-1}\log\Lambda, \quad k \geq 1. 
    \label{Vk-eq}
\end{align}
Since $V_0$ is assumed to be invertible, see (\ref{V0Vk}), 
we can rewrite (\ref{V0-eq}) and (\ref{Vk-eq}) as 
\begin{align}
  [\Lambda^a,V_0] + HV_0 &= 0, \label{V0-eq2}\\
  [\Lambda^a,V_0^{-1}V_k] &= V_0^{-1}[\log\Lambda,V_{k-1}]. 
    \quad k \geq 1. \label{Vk-eq2}
\end{align}

\begin{prop}
There are difference operators of the form (\ref{V0Vk}) 
with polynomial coefficients $v_{kn} = v_{kn}(s)$ 
that satisfy (\ref{V0-eq2}) and (\ref{Vk-eq2}). 
\end{prop}

\proof
We first solve (\ref{V0-eq2}). This equation 
can be translated to the difference equations 
\beq
  v_{0,n+a}(s+a) - v_{0,n+a}(s) = - Hv_{0n}(s) 
  \label{v0a-deq}
\eeq
for the coefficients $v_{0n}$ of $V_0$. 
Starting from $v_{00} = 1$ and $v_{01} = \cdots = v_{0,a-1} = 0$, 
we can find the $v_{0n}$'s recursively 
with the aid of the difference identity 
\begin{gather}
  (s+a)s(s-a)\cdots(s-(k-2)a) - s(s-a)\cdots(s-(k-1)a)
  \notag\\
  = kas(s-a)\cdots(s-(k-2)a) 
  \label{factorial}
\end{gather}
among the $a$-step factorial products as follows. 
Let us examine the difference equation 
\[
  v_{0a}(s+a) - v_{0a}(s) = - Hv_{00}(s) = - s + 1/2
\]
at the first stage of the recursion.  
The identity (\ref{factorial}) for $k = 2$ and $k = 1$ 
gives 
\[
  (s+a)s - s(s-a) = 2as,\quad (s+a)s - s = a. 
\]
Hence a polynomial solution of the difference equation 
can be obtained in the form 
\[
  v_{0a}(s) = - \frac{1}{2a}(s+a)s + \frac{1}{2a}s.
\]
Since $v_{01} = \cdots = v_{0,a-1} = 0$, 
the subsequent $a$ terms $v_{0,a+1},\cdots,v_{0,2a-1}$ 
can be chosen to be equal to $0$.  
The next non-trivial stage is the difference equation 
\[
  v_{0,2a}(s+a) - v_{0,2a}(s) = - Hv_{0a}(s). 
\]
Expanding $-Hv_{0a}(s)$ into a linear combination 
of $s(s-a)(s-2a)$, $s(s-a)$ and $s$, 
we can apply the identity (\ref{factorial}) 
for $k = 2,1,0$ to find a polynomial solution 
of this equation.  Repeating this procedure, 
we obtain a set of polynomials $v_{0a} $ 
that satisfy (\ref{v0a-deq}). 
We now turn to (\ref{Vk-eq2}) and solve these equations 
step-by-step with respect to $k$.  
Suppose that $V_{k-1}$ has been constructed to be 
a difference operator of the form (\ref{V0Vk}) 
with polynomial coefficients.  (\ref{Vk-eq2}) consists 
of the difference equations 
\[
  v'_{k,n+a}(s+a) - v'_{k,n+a}(s) = f_{kn}(s) 
\]
for the coefficients of 
\[
  V_0^{-1}V_k = \sum_{n=ka}^\infty v'_{kn}\Lambda^{-n},\quad
  V_0^{-1}[\log\Lambda,V_{k-1}] = \sum_{n=(k-1)a}^\infty f_{kn}\Lambda^{-n}.
\]
Since $f_{kn}$ is a polynomial in $s$, we can find 
a polynomial $v'_{k,n+a}$ that satisfies this difference equation. 
Thus $V_0^{-1}V_k$, hence $V_k$, becomes a difference operator 
with polynomial coefficients. 
\qed

\begin{remark}
The intertwining relations (\ref{V-rel}) and (\ref{Vbar-rel}) 
do not determine $V$ and $\bar{V}$ uniquely, leaving 
the gauge freedom 
\[
  V \to V\left(1 + \sum_{n=1}^\infty c_n\Lambda^{-n}\right),\quad 
  \bar{V} \to \left(1 + \sum_{n=1}^\infty\bar{c}_n\Lambda^n\right)\bar{V}
\]
of multiplying difference operators with constant coefficients 
$c_n,\bar{c}_n$. 
\end{remark}

\begin{remark}
Since the $V_k$'s take the particular form as shown in (\ref{V0Vk}), 
$V$ itself can be expressed as 
\[
  V = 1 + \sum_{n=1}^\infty\sum_{k=0}^\infty\nu^kv_{kn}\Lambda^{-n}. 
\]
The sum over $k$ is actually a finite sum because $v_{kn} = 0$ 
for $k > n/a$.  Therefore the coefficients $v_n$ of $V$ 
are polynomials in both $s$ and $\nu$.  Thus there is 
no divergence problem for the expansion into powers of $\nu$.  
This is also the case for $\bar{V}$.  
\end{remark}

\section{Non-equivariant limit}

The leading terms $V_0,\bar{V}_0$ in the expansion 
(\ref{nu-expansion}) may be thought of as 
the non-equivariant limit
\[
  V_0 = \lim_{\nu\to 0}V,\quad 
  \bar{V}_0 = \lim_{\nu\to 0}\bar{V} 
\]
of $V,\bar{V}$.  These operators will play a role 
in the non-equivariant Gromov-Witten theory of $\CC\PP^1$.
Similar ideas can be found in the recent papers 
of Chen and Guo \cite{CG19} and Alexandrov \cite{Alexandrov20}.  
Let us consider this issue briefly. 

In the non-equivariant limit, the operator $U$ 
in the factorization problem (\ref{factorization}) 
is replaced by 
\beq
  U_0 = V_0^{-1}e^{\Lambda^a/a}Q^He^{\Lambda^{-b}/b}\bar{V}_0^{-1}.
  \label{U0}
\eeq
$V_0$ and $\bar{V}_0$ satisfy the intertwining relations 
\beq
  (\Lambda^a + H)V_0 = V_0\Lambda^a,\quad 
  \bar{V}_0(\Lambda^{-b} + H) = \Lambda^{-b}\bar{V}_0. 
\eeq
(\ref{..U=U..}) turns into 
\beq
  \Lambda^aU_0 = U_0\Lambda^{-b},
  \label{..U0=U0..}
\eeq
which implies that $L$ and $\bar{L}$ satisfy 
the reduction condition
\beq
  L^a = \bar{L}^{-b}
  \label{LLbar-rel0}
\eeq
to the bigraded Toda hierarchy of type $(a,b)$ 
\cite{Carlet06}. 

Actually, we have the refinement 
\beq
  \Lambda^a U_0 
  = U_0\Lambda^{-b}
  = V_0^{-1}e^{\Lambda^a/a}HQ^He^{\Lambda^{-b}/b}\bar{V}_0^{-1}
  \label{U0-rel2}
\eeq
of (\ref{..U0=U0..}), which can be derived 
in the same way as in the derivation of (\ref{..U=U..}). 
Exponentiating (\ref{U0-rel2})  gives 
\begin{align}
  &\exp\left(\sum_{k=1}^\infty T_k\Lambda^{ka}\right)U_0 
  = U_0\exp\left(\sum_{k=1}^\infty T_k\Lambda^{-kb}\right) \notag\\
  &= V_0^{-1}e^{\Lambda^a/a}\exp\left(\sum_{k=1}^\infty T_kH^k\right)
     Q^He^{\Lambda^{-b}/b}\bar{V}_0^{-1}. 
  \label{U0-rel}
\end{align}
The new variables $T_k$ can be identified 
with coupling constants of the non-equivariant 
Gromov-Witten theory \cite{OP02a,CG19,Alexandrov20}. 
In the language of fermions, (\ref{U0-rel}) 
amount to relations of the form 
\begin{align}
  &\langle s|\exp\left(\sum_{k=1}^\infty T_kJ_{ka}\right)g_0|s\rangle 
   = \langle s|g_0\exp\left(\sum_{k=1}^\infty T_kJ_{-kb}\right)|s\rangle
   \notag \\
  &= \langle s|e^{J_a/a}\exp\left(\sum_{k=1}^\infty T_kP_k\right)
     Q^He^{J_{-b}/b}|s\rangle, 
   \label{g0-rel}
\end{align}
where 
\beq
  g_0 = V_0^{-1}e^{J_a/a}Q^He^{J_{-b}/b}\bar{V}_0^{-1}
  \label{g0}
\eeq
and $P_k$'s are fermion operators that correspond to $H^k$ 
by the correspondence (\ref{fermion-dop}).  
Note that $V_0^{-1}$ and $\bar{V}_0^{-1}$ disappear 
in the last line of (\ref{g0-rel}) because 
$\langle s|V_0^{-1} = \langle s|$ 
and $\bar{V}_0^{-1}|s\rangle = |s\rangle$. 
Thus the coupling constants $T_k$ can be identified 
with part of the time variables of the bigraded Toda hierarchy. 

Let us mention that this construction can capture 
the extended (logarithmic) flows of the 1D/bigraded 
Toda hierartchy \cite{CDZ04,Carlet06} as well. 
Let $\tilde{\bsT} = \{\tilde{T}_k\}_{k=1}^\infty$ 
be time variables of the extended flows and deform $U_0$ as 
\beq
  U_0(\tilde{\bsT}) 
  = \exp\left(\sum_{k=1}^\infty\tilde{T}_k\Lambda^{ka}\log\Lambda\right)
    U_0\exp\left(- \sum_{k=1}^\infty
      \tilde{T}_k\Lambda^{-kb}\log\Lambda\right).
\eeq
By the intertwining relation (\ref{..U0=U0..}) of $U_0$, 
the deformed operator $U_0(\tilde{\bsT})$ 
satisfies the differential equations 
\beq
  \frac{\rd U_0(\tilde{\bsT})}{\rd\tilde{T}_k} 
  = \Lambda^{ka}[\log\Lambda,U_0(\tilde{\bsT})] 
  = \Lambda^{ka}\frac{\rd U_0(\tilde{\bsT})}{\rd s}
\eeq
hence turns out to be a genuine difference operator 
(i.e., does not contain $\log\Lambda$). Thus 
the factorization problem (\ref{factorization}) 
persists to be meaningful.  
The associated dressing operators $W,\bar{W}$ 
satisfy the Sato equations 
\[
  \frac{\rd W}{\rd\tilde{T}_k} 
   = C_kW - W\Lambda^{ka}\log\Lambda,\quad 
  \frac{\rd\bar{W}}{\rd\tilde{T}_k} 
   = C_k\bar{W} - \bar{W}\Lambda^{-kb}\log\Lambda 
\]
of the extended flows \cite{CDZ04,Carlet06}. 
The $C_k$'s are defined as 
\[
  C_k = \calL^k\log\Lambda 
    - (W\Lambda^{ka}W^{-1}\frac{\rd W}{\rd s}W^{-1})_{\ge 0}
    - (\bar{W}\Lambda^{-kb}\bar{W}^{-1}\frac{\rd\bar{W}}{\rd s}\bar{W}^{-1})_{<0}, 
\]
where $\calL$ denotes the difference operator 
of finite order defined by both sides 
of the reduction condition (\ref{LLbar-rel0}).

\section{Conclusion}

We have reformulated Okounkov and Pandharipande's 
dressing operators as difference operators 
that satisfy the intertwining relations (\ref{V-rel}) 
and (\ref{Vbar-rel}).  This formulation fits well 
into the Lax formalism of the 2D Toda hierarchy.  
These dressing operators are building blocks 
of the factorization problem (\ref{factorization}) 
that captures Okounkov and Pandharipande's tau function 
in the Lax formalism.  It is a rather immediate consequence 
of the factorization problem that the Lax operators 
satisfy the reduction condition (\ref{LLbar-rel}) 
to the equivariant Toda hierarchy.  
We have thus found a new explanation to the question 
of why the equivariant Toda hierarchy emerges 
in the Gromov-Witten theory of $\CC\PP^1$.

\subsection*{Acknowledgements}

This work was partly supported by the JSPS Kakenhi Grant 
JP18K03350.


\begin{thebibliography}{99}

\bibitem{EY94}
T.~Eguchi and S.-K.~Yang, 
The topological $CP^1$ model and the large-$N$ matrix integral, 
Mod. Phys. Lett. {\bf A9} (1994), 2893--2902.

\bibitem{EHY95}
T.~Eguchi, K.~Hori, and S.-K.~Yang, 
Topological $\sigma$ models and large-$N$ matrix integral, 
Internat. J. Modern Phys. {\bf A10} (1995), 4203--4224.

\bibitem{Pandharipande99}
R.~Pandharipande, 
The Toda equations and the Gromov-Witten theory 
of the Riemann sphere, 
Lett. Math. Phys. {\bf 53} (2000), 59--74. 

\bibitem{Okounkov00}
A.~Okounkov, 
Toda equations for Hurwitz numbers, 
Math. Res. Lett. {\bf 7} (2000), 447--453. 

\bibitem{Getzler01}
E.~Getzler, 
The Toda conjecture,
K.~Fukaya et al. (eds.), 
\textit{Symplectic Geometry and Mirror Symmetry}, 
World Scientific, 2001, pp. 51--79. 

\bibitem{OP02a}
A.~Okounkov and R.~Pandharipande, 
Gromov-Witten theory, Hurwitz theory, and completed cycles, 
Ann. Math. {\bf 163} (2006), 517--560. 

\bibitem{OP02b}
A.~Okounkov and R.~Pandharipande, 
The equivariant Gromov–Witten theory of $\mathbf{P}^1$, 
Ann. Math. {\bf 163} (2006), 561--605. 

\bibitem{Zhang02}
Y. Zhang, 
On the $CP^1$ topological sigma model 
and the Toda lattice hierarchy, 
J. Geom. Phys. {\bf 40} (2002), 215--232.

\bibitem{DZ03}
B.~Dubrovin and Y.~Zhang, 
Virasoro symmetries of the extended Toda hierarchy, 
Comm. Math. Phys. {\bf 250} (2004), 161--193. 

\bibitem{Milanov0605}
T.~Milanov, 
Gromov-Witten theory of $\CC\PP^1$ and integrable hierarchies, 
arXiv:math-ph/0605001.

\bibitem{MT0607}
T.~Milanov and H.-H.~Tseng, 
The spaces of Laurent polynomials, Gromov-Witten theory 
of $\PP^1$-orbifolds, and integrable hierarchies, 
J. Reine Angew. Math. {\bf 622} (2008), 189--235.

\bibitem{CvdL13}
G.~Carlet and J.~van~de~Leur, 
Hirota equations for the extended bigraded Toda hierarchy 
and the total descendant potential of $\CC\PP^1$ orbifolds, 
J. Phys. A: Math. Theor. {\bf 46} (2013), 405205.

\bibitem{MST14}
T.~Milanov, Y.~Shen and H.-H.~Tseng, 
Gromov-Witten theory of Fano orbifold curves, 
Gamma integral structures and ADE-Toda hierarchies, 
Geom. Topol. {\bf 20} (2016), 2135--2218. 

\bibitem{CM1910}
J.~Cheng and T.~Milaov, 
Gromov-Witten invariants and the extended D-Toda hierarchy, 
arXiv:1909.12735. 

\bibitem{CDZ04}
G. Carlet, B. Dubrovin and Y. Zhang, 
The extended Toda hierarchy,
Moscow Math. J. {\bf 4} (2004), 313-332, 534. 

\bibitem{Carlet06}
G.~Carlet, 
The extended bigraded Toda hierarchy, 
J. Phys. A: Math. Gen. {\bf 39} (2006), 9411--9435. 

\bibitem{CM1909}
J.~Cheng and T.~Milaov, 
The extended D-Toda hierarchy, 
arXiv:1909.12735. 

\bibitem{Milanov0508}
T.~E.~Milanov, 
The equivariant Gromov-Witten theory of $\CC\PP^1$ 
and integrable hierarchies, 
Int. Math. Res. Not. (2008), rnn 073, 21 pp. 

\bibitem{MT0707}
T.~Milanov and H.~H.~Tseng, 
Equivariant orbifold structures on the projective line 
and integrable hierarchies, 
Adv. Math. {\bf 226} (2011), 641--672. 

\bibitem{Johnson09}
P.~D.~Johnson, 
Equivariant Gromov-Witten theory of one dimensional stacks, 
Comm. Math. Phys. {\bf 327} (2014), 333--386. 

\bibitem{Getzler04}
E.~Getzler,
The equivariant Toda lattice, 
Publ. RIMS, Kyoto University, {\bf 40} (2004), 507--534. 

\bibitem{Givental01}
A.~Givental, 
Gromov-Witten invariants and quantization 
of quadratic Hamiltonians, 
Moscow Math. J. {\bf 1} (2001), 551--568.

\bibitem{Takasaki18}
K.~Takasaki,
Toda hierarchies and their applications, 
J. Phys. A: Math. Theor. {\bf 51} (2018), 203001 (35pp). 

\bibitem{TT95}
K.~Takasaki and T.~Takebe, 
Integrable hierarchies and dispersionless limit, 
Rev. Math. Phys. {\bf 7} (1995), 743--808. 

\bibitem{CG19}
C.-Y.~Chen and S.~Guo, 
Quantum curve and bilinear fermionic form 
for the orbifold Gromov-Witten theory of $\PP[r]$, 
arXiv:1912.00558. 

\bibitem{Alexandrov20}
A.~Alexandrov, 
Matrix model for the stationary sector of Gromov-Witten theory 
of $\mathbf{P}^1$, 
arXiv:2001.08556. 

\end{thebibliography}
\end{document}